\documentclass[11pt]{article} % Font size (can be 10pt, 11pt or 12pt) and paper size (remove a4paper for US letter paper)

\usepackage[T1]{fontenc}
\usepackage{mathpazo}
\usepackage[scaled=.85]{beramono} % code font
\usepackage{graphicx} % Required for including pictures
\usepackage{amsmath}
\usepackage{amsfonts}
\usepackage{multirow}
\usepackage{booktabs}
\usepackage{array}
\usepackage{bm}
\usepackage{xcolor}
\usepackage[left=2.25cm,right=2.25cm,bottom=2.5cm,top=2.5cm]{geometry}
\usepackage{parskip}
\usepackage{setspace}
\usepackage[unicode=true,bookmarks=false,pdfstartview={FitH},colorlinks=true,citecolor=blue,linkcolor=blue,urlcolor=blue]{hyperref}
\usepackage[justification=centering,font=small]{caption}
\usepackage{natbib}
\usepackage{xspace}
\graphicspath{{figures/}}
\usepackage{algorithm}
\usepackage{algpseudocode}
\algblockdefx[RFunction]{RFunction}{EndRFunction}
[2]{\textbf{function} \textsc{#1}\,(#2)}
[1][]{\textbf{return}\,#1}

\makeatletter
\renewcommand\@biblabel[1]{\textbf{#1.}} % Change the square brackets for each bibliography item from '[1]' to '1.'
\renewcommand{\@listI}{\itemsep=0pt} % Reduce the space between items in the itemize and enumerate environments and the bibliography
\renewcommand{\maketitle}{ % Customize the title - do not edit title and author name here, see the TITLE block below

  \begin{center} % Right align
  {\bfseries\LARGE\@title} % Increase the font size of the title

  \vspace{2em} % Some vertical space between the title and author name

  {\bfseries\large\@author} % Author name

  \vspace{.75em}
  
  {\bfseries\@date} % Date

  \vspace{1.5em} % Some vertical space between the author block and abstract

  \end{center}
}

\newcommand{\Mo}{\mathcal{M}}
\newcommand{\St}{\mathcal{S}}
\newcommand{\R}{\texttt{R}\xspace}
\newcommand{\s}{\sigma}
\newcommand{\g}{\gamma}
\newcommand{\dd}{\delta}
\newcommand{\veps}{\varepsilon}
\newcommand{\XX}{\bm{X}}
\newcommand{\WW}{\bm{W}}
\newcommand{\ZZ}{\bm{Z}}
\newcommand{\RR}{\bm{R}}
\renewcommand{\SS}{\bm{S}}
\newcommand{\UU}{\bm{U}}
\newcommand{\YY}{\bm{Y}}
\newcommand{\xx}{\bm{x}}
\newcommand{\ww}{\bm{w}}
\newcommand{\zz}{\bm{z}}
\newcommand{\bbe}{\bm{\beta}}
\newcommand{\gga}{\bm{\gamma}}
\newcommand{\tth}{\bm{\theta}}
\newcommand{\dde}{\bm{\delta}}
\newcommand{\ssi}{\bm{\sigma}}
\newcommand{\OOm}{\bm{\Omega}}
\newcommand{\DD}{{\bm{\mathcal{D}}}}
\DeclareMathOperator{\N}{\mathcal N}
\DeclareMathOperator{\Unif}{Uniform}
\newcommand{\Cmax}{C_{\tx{max}}}
\newcommand{\tx}[1]{\textrm{#1}}
\renewcommand{\tt}[1]{\texttt{#1}}
\newcommand{\ind}{\stackrel {\textrm{ind}}{\sim}}
\newcommand{\rv}[3][1]{#2_{#1},\ldots,#2_{#3}}
\DeclareMathOperator*{\argmax}{arg\,max}
\newcommand{\where}{\quad\tx{where}\quad}
\renewcommand{\sp}[1]{^{(#1)}}
\newcommand{\lp}[1]{_{(#1)}}
\newcommand{\diag}{\textrm{diag}}
\DeclareMathOperator{\se}{se}
\DeclareMathOperator{\sd}{sd}
\DeclareMathOperator{\var}{var}

\newcommand{\cv}{\tx{CV}}
\newcommand{\rng}{\tx{RNG}}
\newcommand{\rsq}{R^2}
\newcommand{\rmse}{\tx{RMSE}}

\newcommand{\sdot}{\raisebox{.1em}{\textrm{\tiny$\bullet$}}}
\def\bsqrt{\mathpalette\DHLhksqrt}
\def\DHLhksqrt#1#2{%
\setbox0=\hbox{$#1\sqrt{#2\,}$}\dimen0=\ht0
\advance\dimen0-0.2\ht0
\setbox2=\hbox{\vrule height\ht0 depth -\dimen0}%
{\box0\lower0.4pt\box2}}

\newcommand{\nserc}{Natural Sciences and Engineering Research Council of Canada}
\newcommand{\grantnumber}{RGPIN-2014-04225}
\title{A Heteroscedastic Accelerated Failure Time Model\\ for Survival Analysis\\ % Title
} % Subtitle

\author{Yifan Wang$^\star$ \quad Tian You$^\star$ \quad Martin Lysy$^\dagger$  % Author% Author
  \\
  \vspace{.75em}
  University of Waterloo} % Institution

\date{%\vspace{.5in}
  {\today}} % Date

\begin{document}

{\let\thefootnote\relax\footnotetext{\hspace{-2.2em}$^\star$Equally contributing authors.}}
{\let\thefootnote\relax\footnotetext{\hspace{-2.2em}$^\dagger$Corresponding author: Dr.~Martin Lysy $\sdot$ Department of Statistics \& Actuarial Science $\sdot$ University of Waterloo\\
  200 University Avenue West $\sdot$ Waterloo ON $\sdot$ N2L 3G1 $\sdot$  Tel.: (519) 888-4567 x35503 $\sdot$  Email: mlysy@uwaterloo.ca\\ The authors gratefully acknowledge funding from the \nserc{} Discovery Grant \grantnumber.}}

\maketitle % Print the title section

% ----------------------------------------------------------------------------------------
%	ABSTRACT AND KEYWORDS
% ----------------------------------------------------------------------------------------

% \renewcommand{\abstractname}{Summary} % Uncomment to change the name of the abstract to something else

\begin{abstract}
  \noindent Nonparametric and semiparametric methods are commonly used in survival analysis to mitigate the bias due to model misspecification.
  However, such methods often cannot estimate upper-tail survival quantiles when a sizable proportion of the data are censored, in which case parametric likelihood-based estimators present a viable alternative.
  % However, such methods often can only estimate lower survival quantiles when a sizeable proportion of the data are censored.
  % In such situations, parametric likelihood-based estimators present a viable alternative when upper-quantile analyses are desired.
  In this article, we extend a popular family of parametric survival models which make the Accelerated Failure Time (AFT) assumption to account for heteroscedasticity in the survival times.
  The conditional variances can depend on arbitrary covariates, thus adding considerable flexibility to the homoscedastic model. 
  % The heteroscedasticity is allowed to depend on arbitrary covariates, thereby adding substantial flexibility to the homoscedastic AFT model.
  We present an Expectation-Conditional-Maximization (ECM) algorithm to efficiently compute the HAFT maximum likelihood estimator with right-censored data.  The methodology is applied to the heavily censored data from a colon cancer clinical trial, for which a new type of highly stringent model residuals is proposed.  Based on these, %heteroscedastic modeling
  % Based on a novel, highly stringent definition of model residuals for heavily censored data, 
  % Applying the methodology to the analysis of a colon cancer clinial trial,
  % a stringent residual-based analysis indicates th
  % In an application to the analysis of a colon cancer clinical trial,
  % a stringent residual-based evaluation 
  % An Expectation-Conditional-Maximization (ECM) algorithm for maximum likelihood with right-censored data is presented.  In an application to the analysis of a colon cancer study,
  % heteroscedastic modeling was found to considerably improve the quality of fit relative to its hop
  the HAFT model was found to eliminate most outliers from its homoscedastic counterpart.

  \smallskip
  
%  heteroscedastic modeling was found to eliminate most outliers, and predict better response to treatment in the upper-tail quantiles. \\
  % to slightly decrease the average size of prediction intervals. \\

  % This adds substantial modeling flexibility, and we show how to easily and rapidly compute maximum likelihood estimators for the proposed model in the presence of censoring.
  
  % \noindent{}While the Cox Proportional Hazard model is a fundamental tool in survival analysis, its semi-parametric nature precludes the estimation of upper survival quantiles in the presence of heavy censoring.  In contrast, fully parametric models do not suffer from this issue -- at the expense of additional modeling assumptions.  In this article, we extend a popular family of parametric models which make the Accelerated Failure Time (AFT) assumption to account for heteroscedasticity in the log-survival times.  This adds substantial modeling flexibility, and we show how to easily and rapidly compute maximum likelihood estimators for the proposed model in the presence of censoring.  In an application to the analysis of a colon cancer study, we found that heteroscedastic modeling greatly diminished the significance of outliers, while even slightly decreasing the average size of prediction intervals. \\

  % \noindent \textit{Keywords:} Accelerated Failure Time assumption, Heteroscedastic modeling, Right-censored lifetimes, Expectation-Conditional Maximization % Keywords

\end{abstract}

% \vspace{20pt} % Some vertical space between the abstract and first section

% ----------------------------------------------------------------------------------------
%	ESSAY BODY
% ----------------------------------------------------------------------------------------

\section{Introduction}

% TODO:
% \begin{itemize}
% \item Shift focus to NP = CPH + QR
% \item Add example with QR
% \item Add table with covariate definitions
% \end{itemize}

When modeling the dependence of survival times $T$ on a set of predictors $\XX = (\rv X D)$, nonparametric and semiparametric estimators are often used in medical applications to mitigate the adverse effects of model misspecification.  Some of the most well-known estimators of this type are based on the Cox Proportional Hazards (CPH) model~\citep{cox72}.  The CPH model is highly flexible, straightforward to fit, and accommodates right-censored failure times -- a ubiquitous feature of medical lifetime data.  Another popular class of semiparametric survival estimators are those of quantile regression (QR) models and their censoring extensions~\citep{koenker.bassett78,koenker05,powell86,portnoy03,peng.huang08}.
However, both QR and CPH semiparametric models produce truncated estimators of the conditional survival function
% can fail to produce estimates of the conditional survival function
\[
  S(t \mid \xx) = \Pr(T > t \mid \XX = \xx)
\]
when the largest survival times in the dataset are censored~\citep[e.g.,][]{moeschberger.klein85,peng.huang08,koenker18}.  That is, the CPH and QR estimators are of the form $\hat S(t \mid \xx) \equiv 1-\alpha_0$ beyond a data-dependent threshold $t > t_0$, such that the corresponding quantile estimator
\[
  \hat Q(\alpha \mid \xx) = \hat S^{-1}(1-\alpha \mid \xx)
\]
for $\alpha > \alpha_0$ is undefined.  This can become a serious limitation when the censoring rate is high~\citep[e.g.,][]{sy.taylor00}.
% However, both QR and CPH semiparametric models produce truncated estimators of the conditional survival function
% % can fail to produce estimates of the conditional survival function
% \[
%   S(t \mid \xx) = \Pr(T > t \mid \XX = \xx),
% \]
% such that $\hat S(t \mid \xx) \equiv 1-\alpha_0$ beyond a data-dependent threshold $t > t_0$, when the largest survival times in the dataset are censored~\citep[e.g.,][]{moeschberger.klein85,peng.huang08,koenker18}.  Consequently, the corresponding quantile estimator
% \[
%   \hat Q(\alpha \mid \xx) = \hat S^{-1}(1-\alpha \mid \xx)
% \]
% is undefined for $\alpha > \alpha_0$, which becomes an important concern when the censoring rate is high~\citep[e.g.,][]{sy.taylor00}.
In contrast, parametric likelihood-based estimators do not suffer from this issue, thus presenting a viable alternative in heavy censoring situations when analyses of upper-tail survival quantiles are desired.

A popular parametric-likelihood approach to conditional quantile estimation operates under the Accelerated Failure Time (AFT) assumption~\citep{wei92,kalbfleisch.prentice02}; namely, that the conditional survival time is given by
\[
  \log(T) = \mu(\XX) + \veps,
\]
where $\veps \sim f_0(t) $ is a random variable which does not depend on $\XX$.  AFT models have an appealing interpretation for quantile estimation: the relation between the conditional survival function $S(t\mid\xx)$ and the baseline survival function $S_0(t)$ of $\veps$ is
\[
  S(t\mid \xx) = S_0(\lambda(\xx) \cdot t), \where \lambda(\xx) = e^{-\mu(\xx)}.
\]
However, as with any parametric model, incorrect specification of scale and distribution functions $\mu(\xx)$ and $f_0(t)$ can adversely affect inferential results.

The purpose of this article is to relax the homoscedasticity assumption made by the AFT model on the conditional log-survival distribution.  Much attention has been devoted to this in the context of random individual-level effects, referred to in the literature as ``frailty modeling''~\citep[e.g.][]{hougaard91, keiding.etal97, pan01, zhang.peng09}.  We adopt instead a covariate-dependent heteroscedastic modeling approach of the form
\begin{equation}\label{eq:haft}
  \log(T) = \mu(\XX) + \sigma(\XX) \cdot \veps.
\end{equation}
Estimation for location-scale type regression models such as~\eqref{eq:haft} has been studied extensively; see e.g.,~\cite{muller.stadtmuller87, cai.wang08} and e.g.,~\cite{hsieh96, zeng.lin07, zhang.davidian08, su.etal12} for nonparametric and semiparametric approaches, respectively.  With certain restrictions,~\eqref{eq:haft} can also be viewed as a quantile regression model~\citep[e.g.][]{koenker.machado99}.

Fully parametric likelihoods under model~\eqref{eq:haft} have been studied by e.g.,~\cite{boscardin.gelman96,smyth02}.  Following these authors, we consider the generalized linear regression-type model specification
\begin{equation}\label{eq:hpar}
  \mu(\xx) = \xx'\bbe, \qquad \s^2(\xx) = \exp(\xx'\gga), \qquad \veps \sim \N(0,1).
\end{equation}
% As with its homoscedastic counterpart, the heteroscedastic AFT (HAFT) model~\eqref{eq:hpar} 
% The primary advantage of this particular Heteroscedastic Accelerated Failure Time (HAFT) model~\eqref{eq:hpar} is the availability of an efficient algorithm for computing maximum likelihood estimators of $\bbe$ and $\gga$ in the censoring-free setting~\citep[e.g.,][]{smyth89,verbyla93}.  In this article, we derive an Expectation-Conditional-Maximization (ECM) algorithm~\citep{meng.rubin93} to extend these calculations to the case of right-censored data.
% As a generalization of its homoscedastic counterpart, the HAFT model adds considerable flexibility to the modeling of survival times.  We illustrate this with data from a colon cancer clinical trial exhibiting a high proportion of right-censored observations.  
While the adequacy of any failure time model clearly varies from one dataset to another, we shall advocate here that the Heteroscedastic Accelerated Failure Time (HAFT) model~\eqref{eq:hpar} is an attractive addition to the survival modeling toolkit for a number of reasons:
\begin{itemize}    
\item \textsl{Interpretability.} As with the homoscedastic AFT model, the conditional survival function of the HAFT model can be easily related to the baseline survival function of $\veps$:
  \[
    S(t \mid \xx) = S_0\big(\lambda(\xx) \cdot t^{\alpha(\xx)}\big), \where \begin{array}{rl} \lambda(\xx) & = e^{-\mu(\xx)/\s(\xx)},\\ \alpha(\xx) & = 1/\s(\xx).\end{array}
  \]
  Consequently, it is straightforward to calculate $S(t \mid \xx)$ under~\eqref{eq:hpar} for any combination of $t$ and $\xx$ using the quantile function of a standard normal distribution.
\item \textsl{Tractability.}  A distinct advantage of the specific HAFT formulation~\eqref{eq:haft} is the availability of an efficient algorithm for computing maximum likelihood estimators of $\bbe$ and $\gga$ in the censoring-free setting~\citep[e.g.,][]{smyth89,verbyla93}.  In this article, we derive an Expectation-Conditional-Maximization (ECM) algorithm~\citep{meng.rubin93} to efficiently extend these computations to the right-censoring case.
  % -- full details and an implementation using standard statistical software are provided in Section~\ref{sec:mthd}.  Moreover, confidence intervals for the model parameters and quantile estimates can readily be constructed from the Hessian matrix of the loglikelihood.  

  % \item \textsl{Censoring.}  The HAFT model~\eqref{eq:hpar} admits a simple Expectation-Conditional Maximization (ECM) algorithm~\citep{meng.rubin93} to estimate $\bbe$ and $\gga$ in the presence of right-censored failure times (described in Section~\ref{sec:necm}).
\item \textsl{Flexibility}. Adding conditional heteroscedasticity to the AFT model adds considerable flexibility to the modeling of survival times.  We demonstrate this both with a simulation study indicating that %failure to account for
  even a small degree of unmodeled heteroscedasticity can lead to considerable bias in quantile estimation, and with data from a colon cancer clinical trial exhibiting a high proportion of censored survivals.  To assess goodness-of-fit for these heavily censored data, a new type of highly stringent model residuals is proposed.  Based on these residuals, the HAFT model
  % stringent residual goodness-of-fit analysis is conducted, 
  % The HAFT model
  was found to have far fewer outliers than its homoscedastic counterpart. %, and even to slightly decrease the average size of prediction intervals. 
\end{itemize}
Elaborating on these points, the remainder of this article is organized as follows.  The efficient maximum likelihood estimation algorithm for the HAFT model~\eqref{eq:haft} with right-censored data is presented in Section~\ref{sec:mthd}.  A simulation study comparing AFT and HAFT models for the purpose of quantile estimation is presented in Section~\ref{sec:sim}.  The analysis of the colon cancer data %A comparison of its performance on the colon cancer data relative to the homoscedastic AFT model
is presented in Section~\ref{sec:eda}.  We conclude with a discussion of further work in Section~\ref{sec:disc}.

% ------------------------------------------------

\section{Maximum Likelihood Estimation for the HAFT Model}\label{sec:mthd}

In order to present parameter fitting algorithms for the HAFT model we introduce the following notation. Let $R_i = \log(T_i)$ and $\XX_i = (\rv [i1] X {iD})$ denote the log-survival time and predictors for subject $i$.  For ease of exposition, we write the HAFT model as
\begin{equation}\label{eq:hlm}
  R_i \mid \XX_i \ind \N\Big(\WW_i' \bbe, \exp(\ZZ_i' \gga)\Big),
\end{equation}
where $\WW_i = (\rv [i1] W {ip}) = \bm{F}(\XX_i)$ and $\ZZ_i = (\rv [i1] Z {iq}) = \bm{G}(\XX_i)$.  The model parameters are $\bbe = (\rv \beta p)$ and $\gga = (\rv \g q)$, and the loglikelihood function is
\begin{equation}\label{eq:hll}
  \ell(\bbe,\gga\mid \RR, \XX) = -\frac{1}{2}\sum_{i=1}^{n} \left[\frac{(R_i-\WW_i'\bbe)^2}{\exp(\ZZ_i'\gga)}+\ZZ_i'\gga\right],
\end{equation}
where $\RR = (\rv R n)$ and $\XX = (\rv {\XX} n)$.

\subsection{Estimation Without Censoring}\label{sec:wocens}

We first present a method of calculating the maximum likelihood estimator (MLE) of $\tth = (\bbe, \gga)$ for complete (uncensored) data.  For fixed $\gga$, the conditional loglikelihood for the mean parameters is
\[
  \ell(\bbe\mid \gga, \RR, \XX) = -\frac{1}{2}\sum_{i=1}^{n} \left[\frac{(R_i-\WW_i'\bbe)^2}{\s_i^2}\right], \where \s_i^2 = \exp(\ZZ_i'\gga).
\]
This is the loglikelihood function of a linear model with normal errors and known variances $\s_i^2$.  With $\WW_{n\times p} = \big[\WW_1 \mid \cdots \mid \WW_n\big]'$, it is maximized at
\begin{equation}\label{eq:betaup}
  \hat {\bbe} = (\WW'\OOm \WW)^{-1} {\WW}' \OOm \RR, \where \OOm^{-1} = \diag\big(\rv {\s^2} n \big).
\end{equation}

For fixed $\bbe$, the conditional loglikelihood of the variance parameters is
\begin{equation}\label{eq:gammaglm}
  \ell(\gga\mid \bbe, \RR, \XX) = -\frac{1}{2}\sum_{i=1}^{n} \left[\frac{U_i}{\exp(\ZZ_i'\gga)} + \ZZ_i'\gga\right], \where U_i = (R_i - \WW_i'\bbe)^2.
\end{equation}
This can be recognized as the loglikelihood of a Generalized Linear Model (GLM) for a Gamma distribution with logarithmic link function~\citep[e.g.,][]{nelder.pergibon87,smyth89}.
The latter provides a Fisher scoring algorithm which iteratively updates $\bbe$ and $\gga$ and converges to the MLE~\citep{smyth89}.  While further accelerations are possible~\citep[e.g.,][]{smyth02}, the maximization of GLM likelihoods can be readily accomplished with tools from standard regression software. 
For example, with $\UU_{n \times 1} = (\rv U n)$ and \mbox{$\ZZ_{n\times q} = \big[Z_1 \mid \cdots \mid Z_n\big]$}, the maximizer $\hat {\gga}$ of~\eqref{eq:gammaglm} can be computed in \R with the command
\begin{equation}\label{eq:gammaup}
  \tt{glm($\UU \sim \ZZ$ - 1, family = Gamma("log"))}.
\end{equation}
Our numerical experiments indicate that alternating between maximization of~\eqref{eq:betaup} and of~\eqref{eq:gammaup} converges very quickly to the MLE of $\tth = (\bbe, \gga)$.

\subsection{An ECM Algorithm for Censored Observations}\label{sec:necm}

In the presence of right-censoring, instead of observing the actual log-failure time $R_i$, we observe $Y_i = \min(R_i, C_i)$, where $C_i$ is the censoring time.  We also observe $\delta_i = \mathfrak 1\{R_i < C_i\}$, a binary variable indicating whether or not the survival time of subject $i$ is censored ($\delta_i = 1$ means uncensored).  Assuming that $R$ and $C$ are conditionally independent given $\XX$, the loglikelihood function given the censored data $\YY = (\rv Y n)$ and $\dde = (\rv \dd n)$ is
\begin{equation}\label{eq:hllc}
  \ell(\bbe, \gga \mid \YY, \dde) = \sum_{i=1}^{n} \delta_i \cdot \left[\frac{-(Y_i-\WW_i'\bbe)^2}{2\cdot\exp(\ZZ_i'\gga)}- \frac {\ZZ_i'\gga}{2}\right] +  (1-\delta_i) \cdot \left[1 - \Phi\left(\frac{Y_i - \WW_i'\bbe}{\exp(\ZZ_i'\gga/2)}\right)\right],
\end{equation}
where $\Phi(\textnormal{\scriptsize$\mathcal Z$})$ is the cumulative distribution function (CDF) of the standard normal $\N(0,1)$ distribution.  While~\eqref{eq:hllc} cannot be maximized directly, we describe here an Expectation-Conditional-Maximization (ECM) algorithm~\citep{meng.rubin93} which combines the efficient maximum likelihood calculations of Section~\ref{sec:wocens} with an Expectation-Maximization algorithm for the censored homoscedastic linear model~\citep[e.g.,][]{aitkin81}.% to our heteroscedastic setting.

Let $\bbe\sp t$ and $\gga \sp t$ denote the parameter values at iteration $t$.  For the E-step, the expecation of the complete data loglikelihood is
\begin{align*}
  Q_t(\bbe, \gga) & = E\big[\ell(\bbe, \gga \mid \RR, \XX) \,\mid\, \YY, \dde, \XX, \bbe\sp t, \gga \sp t\big] \\
                  & = E\left[-\frac 1 2 \sum_{i=1}^n \frac{(R_i - \WW_i' \bbe)^2}{\exp(\ZZ_i' \gga)} - \frac 1 2 \sum_{i=1}^n \ZZ_i' \gga \,\,\Big|\,\, \YY, \dde, \XX, \bbe\sp t, \gga \sp t \right] \\
                  & = -\frac 1 2 \sum_{i=1}^n \frac{\tilde S_i\sp t - 2 \tilde R_i\sp t \WW_i'\bbe + (\WW_i' \bbe)^2}{\exp(\ZZ_i' \gga)} - \frac 1 2 \sum_{i=1}^n \ZZ_i' \gga,
\end{align*}
where
\begin{equation}\label{eq:estep}
  \begin{aligned}
    &
    \tilde R_i\sp t
    =
    \begin{cases}
      Y_i & \dd_i = 1 \\
      \s_i\sp t f(\tilde Y_i\sp t) + \mu_i\sp t & \dd_i = 0,
    \end{cases}
    & & & &
    \tilde S_i\sp t
    =
    \begin{cases}
      Y_i^2 & \dd_i = 1 \\
      (\s_i\sp t)^2 g(\tilde Y_i\sp t) + 2 \mu_i\sp t \tilde R_i\sp t & \dd_i = 0,
    \end{cases}
    \\
    &
    \begin{aligned}[t]
      \mu_i\sp t & = \WW_i'\bbe\sp t,
      &  &  &
      \s_i\sp t & = \exp(\tfrac 1 2 \ZZ_i'\gga\sp t),
      \\
      f(a) & = \frac{\varphi(a)}{\Phi(-a)},
      &  &  &
      g(a) & = 1 + \frac{a\varphi(a)}{\Phi(-a)},
    \end{aligned}
    & & & & 
    \tilde Y_i\sp t = \frac{Y_i - \mu_i\sp t}{\s_i\sp t}
  \end{aligned}
\end{equation}
and $\varphi(\textnormal{\scriptsize$\mathcal Z$})$ is the probability and density functions (PDF) of a standard normal distribution, such that for $\mathcal Z \sim \N(0,1)$ we have $f(a) = E[\mathcal Z \mid \mathcal Z > a]$ and \mbox{$g(a) = E[\mathcal Z^2 \mid \mathcal Z > a]$}.

The M-step consists of first obtaining the conditional maximum $\bbe\sp{t+1} = \argmax_{\bbe} Q_t(\bbe, \gga\sp t)$, followed by the conditional maximum $\gga\sp{t+1} = \argmax_{\gga} Q_t(\bbe\sp{t+1},\gga)$.  For the first part, the solution is given by the weighted linear regression estimate
\[
  \bbe\sp{t+1} = (\WW'\OOm\sp t \WW)^{-1} \WW' \OOm \sp t \tilde {\RR}\sp t, \where
  \begin{aligned} (\OOm\sp t)^{-1/2} & = \diag\Big(\s_1\sp t, \ldots, \s_n\sp t \Big), \\
    \tilde {\RR}\sp t & = (\tilde R_1\sp t, \ldots, \tilde R_n\sp t).
  \end{aligned}
\]
For the second part, $\gga\sp{t+1}$ maximizes the objective function
  \[
    Q_t(\bbe\sp{t+1}, \gga) = 
    - \frac 1 2 \left[\sum_{i=1}^n \frac{\tilde U_i\sp{t+1}}{\exp(\ZZ_i'\gga)} + \ZZ_i' \gga \right],
  \]
  where $\tilde U_i\sp{t+1} = \tilde S_i\sp t - 2 \tilde R_i\sp t \WW_i' \bbe\sp{t+1} + (\WW_i' \bbe\sp{t+1})^2$.  Once again this corresponds to the likelihood of the GLM with Gamma response and logarithmic link function, which can be maximized using standard regression software.
% \item \textsl{M-step for $\bbe$}: The conditional maximum $\bbe\sp{t+1} = \argmax_{\bbe} Q_t(\bbe, \gga\sp t)$ is given by the weighted linear regression estimate:
%   \[
%     \bbe\sp{t+1} = (\WW'\OOm\sp t \WW)^{-1} \WW' \OOm \sp t \tilde {\RR}\sp t, \where
%     \begin{aligned} (\OOm\sp t)^{-1/2} & = \diag\Big(\s_1\sp t, \ldots, \s_n\sp t \Big), \\
%       \tilde {\RR}\sp t & = (\tilde R_1\sp t, \ldots, \tilde R_n\sp t).
%     \end{aligned}
%   \]
% \item \textsl{M-step for $\gga$}: Similarly, $\gga\sp{t+1}$ maximizes the objective function
%   \[
%     Q_t(\bbe\sp{t+1}, \gga) = 
%     - \frac 1 2 \left[\sum_{i=1}^n \frac{\tilde U_i\sp{t+1}}{\exp(\ZZ_i'\gga)} + \ZZ_i' \gga \right],
%   \]
%   where $\tilde U_i\sp{t+1} = \tilde S_i\sp t - 2 \tilde R_i\sp t \WW_i' \bbe\sp{t+1} + (\WW_i' \bbe\sp{t+1})^2$.  Once again this corresponds to the likelihood of the GLM with Gamma response and logarithmic link function, which can be maximized using standard regression software.
% \end{itemize}
Alternating between the E-step and each of the conditional M-steps converges to the MLE $\hat {\tth} = (\hat {\bbe}, \hat {\gga})$ of the censored heteroscedastic loglikelihood~\eqref{eq:hllc}.  The exact ECM procedure is described in Algorithm~\ref{alg:ecm}.

We may readily obtain a variance estimator for $\hat \tth$ by calculating
\[
  \widehat{\var}(\hat {\tth}) = -\left[\frac{\partial^2}{\partial \tth^2} \ell(\hat {\bbe}, \hat {\gga} \mid \YY, \dde)\right]^{-1}.
\]
If the objective is to estimate the $\alpha$-level conditional quantile
\[
  Q_\alpha(\xx, \tth) = \exp\big\{\ww'\bbe + \exp(\zz'\gga/2) \cdot \Phi^{-1}(\alpha)\big\},
\]
a natural estimator is
\[
  \hat Q = \exp\big\{\ww\hat{\bbe} + \exp(\zz'\hat{\gga}/2) \cdot \Phi^{-1}(\alpha)\big\},
\]
for which asymptotic theory~\citep[e.g.,][]{oakes77} gives the standard error as
\begin{equation}\label{eq:qse}
  \se(\hat Q) = \sqrt{\bm g(\hat {\tth})' \widehat{\var}(\hat {\tth}) \bm g(\hat {\tth})}, % = (\ww, \tfrac 1 2 \zz \exp(\zz'\gga/2) \cdot \Phi^{-1}(\alpha)).
\end{equation}
where
\[
  \bm g(\tth) = \frac{\partial Q_\alpha(\xx, \tth)}{\partial \tth} = \hat Q \times \big(\ww, \exp(\zz'\gga/2) \cdot \Phi^{-1}(\alpha) \cdot \zz/2\big).
\]
\begin{algorithm}[!htb]
    \small
  \setstretch{1.2}
  \caption{\small ECM algorithm for fitting the HAFT model.  Inputs are the data $\DD = (\YY, \dde, \WW, \ZZ)$, the maximum number of iterations $M$, the error tolerance $\epsilon > 0$, and initial parameter values $(\bbe\sp 0,\gga \sp 0)$.}\label{alg:ecm}
  \begin{algorithmic}
    \RFunction{haftFit}{$\DD, M, \epsilon, \bbe\sp 0, \gga\sp 0$}
    \State $\ell\sp 0 \gets \ell(\bbe\sp 0, \gga \sp 0 \mid \YY, \dde)$ \Comment{Evaluate censored loglikelihood~\eqref{eq:hllc}}
    \For{$t = 0,\ldots,M-1$}
    \State $\big\{\tilde \RR\sp t, \tilde \ssi\sp t, \tilde \SS\sp t\big\} \gets \Call{Estep}{\bbe\sp t, \gga \sp t}$ \Comment{Compute the E-step quantities in~\eqref{eq:estep}}
    \State $\begin{aligned}[m]
      \OOm^{-1/2}\lp t & \gets \diag\Big(\s_1\sp t, \ldots, \s_n\sp t \Big)\\
      \bbe\sp{t+1} & \gets (\WW'\OOm\sp t \WW)^{-1} \WW' \OOm \sp t \tilde {\RR}\sp t
    \end{aligned}$ \Comment{Conditional M-step for $\bbe$}
    \State $\begin{aligned}[m]
      \tilde U_i\sp{t+1} & \gets \tilde S_i\sp t - 2 \tilde R_i\sp t \WW_i' \bbe\sp{t+1} + (\WW_i' \bbe\sp{t+1})^2 \\
      \gga\sp{t+1} & \gets \tt{glm($\UU\sp{t+1} \sim \ZZ$ - 1, family = Gamma("log"))}
    \end{aligned}$ \Comment{Conditional M-step for $\gga$}
    \State $\ell\sp{t+1} \gets \ell(\bbe\sp {t+1}, \gga \sp {t+1} \mid \YY, \dde)$
    \State $\textbf{if}\ \displaystyle \frac{|\ell\sp{t+1} - \ell\sp t|}{0.1 + |\ell\sp{t+1}|} < \epsilon\ \textbf{then}\ \textbf{break}$ \Comment{Terminate early if error tolerance is reached}
    \EndFor
    \EndRFunction $\big\{\bbe\sp{t+1}, \gga\sp{t+1}\big\}$
  \end{algorithmic} \label{alg:gschur}
\end{algorithm}

% \begin{align*}
%   \bm g(\tth) & = \frac{\partial Q_\alpha(\xx, \tth)}{\partial \tth} \\
%               & = \hat Q \times \big(\ww, \exp(\zz'\gga/2) \cdot \Phi^{-1}(\alpha) \cdot \zz/2\big). 
% \end{align*}

% Moreover, a point estim for the conditional $\alpha$-level quantile
% \[
%   w
% \]
% are given by $\hat q = \xx'\bbe + \exp(\zz'\gga/2) \cdot \Phi^{-1}(\alpha)$

\section{Simulation Study}\label{sec:sim}

In order to assess the impact of heteroscedastic modeling on quantile estimation, the following simulation experiment is conducted. 
% in order to assess the added value of heteroscedastic modeling.
The model used to generate survival times is of the form
\begin{equation}
  \label{eq:simmodel}
  \log(T) = R = \beta_0 + \XX'\bbe + \s \exp(\XX'\gga/2) \cdot \veps, \qquad \veps \sim \N(0, 1),
\end{equation}
with $\XX \sim \N(\bm 0, \bm I_p)$, $\beta_j \equiv \beta$, and $\gamma_j \equiv \gamma$. The model used to generate censoring times is $\exp(C) \sim \Unif(0, \Cmax)$.  %We note that under model~\eqref{eq:simmodel}, the true conditional quantile $Q_\alpha(\XX, \tth)$ depends only on $\bar X = \tfrac 1 p \sum_{j=1}^p X_j$.
The value of $\gamma$ is chosen so as to control the coefficient of variation of the conditional log-scale standard deviation,
\[
  \cv = \frac{\bsqrt{\var(\sd(R \mid \XX))}}{E[\sd(R \mid \XX)]} = \bsqrt{\exp(p \gamma^2/4) - 1}.
\]
The values of $\beta$ and $\s$ are chosen within the context of a homoscedastic AFT model
\[
  \log(\tilde T) = \tilde R = \beta_0 + \XX'\bbe + \tilde \sigma \cdot \veps,
\]
where $\tilde \s$ is set to $E[\sd(R \mid X)]$ under the original HAFT model, namely
\[
  \tilde \s = \sigma \exp(p\gamma^2/8) = \sigma A.
\]
Under this hypothetical AFT model, let $\tilde Q_\alpha$ denote the $\alpha$-level (unconditional) quantile of $\tilde T$, such that the range of $\tilde T$ may be controlled by fixing
\[
  \rng = \frac{\tilde Q_{.99}}{\tilde Q_{.01}} = \exp\left\{\big(\Phi^{-1}(.99) - \Phi^{-1}(.01)\big) \cdot \bsqrt{p\beta^2 + \sigma^2 A^2} \right\}.% \approx \exp(4.65 \cdot \bsqrt{p\beta^2 + \sigma^2 A^2}).
\]
Similarly, for a fixed value of $\rng$, the tradeoff between $\beta$ and $\sigma$ may be controlled by fixing the proportion of variance explained in the AFT model,
\[
  \rsq = \frac{\var(E[\tilde R \mid \XX])}{\var(\tilde R)} = \frac{p\beta^2}{p\beta^2 + \sigma^2 A^2}.
\]
% is chosen so as to control the range of the survival times $\tilde T$ under a homoscedastic AFT model $\log(\tilde T) = \XX'\beta + \tilde \s \cdot \veps$, where $\tilde \s$ is set to $E[\sd(R \mid X)]$ under the original HAFT model,
% \[
%   \tilde \s = \sigma \exp(p\gamma^2/8) = \sigma A.
% \]
% That is, let $\tilde Q_\alpha$ denote the $\alpha$-level (unconditional) quantile of $\tilde T$.  Then we control the range of $\tilde T$ by fixing
% \[
%   \rng = \frac{\tilde Q_{.99}}{\tilde Q_{.01}} = \exp\left\{\sigma A \cdot \big(\Phi^{-1}(.99) - \Phi^{-1}(.01)\big)\right\} \approx \exp(\sigma A \cdot 4.65).
% \]
% the coefficient of variation of $T$ under a homoscedastic AFT model $\log(\tilde T) = \XX'\beta + \tilde \s \cdot \veps$, where $\tilde \s$ is set to $E[\sd(R \mid X)]$ under the original HAFT model,
% \[
%   \tilde \s = \sigma \exp(p\gamma^2/8) = \sigma A,
% \]
% such that
% \[
%   \cv_\s = \frac{\sd(\tilde T)}{E[\tilde T]} = \bsqrt{\exp(\sigma^2A^2) - 1}.
% \]
Finally, the range of the censoring variable $\Cmax$ is chosen (via Monte Carlo simulation) 
% is given by $C \sim \Unif(0, \Cmax)$, with $\Cmax$ chosen (via numerical simulation)
so as to control the overall censoring probability $P_C = \Pr(C < T)$.

The simulation experiment was conducted over $B = 1000$ datasets $\DD_b = (\YY\sp b, \dde\sp b, \XX \sp b)$, $b = 1, \ldots, B$, generated for each experimental combination listed in Table~\ref{tab:sim}.
\begin{table}[!htb]
  % \footnotesize
  \centering
  \begin{tabular}{rl}  
    \toprule
    Parameter & Description \\
    \midrule
    $N = 1000$  & Sample size \\
    $p \in \{1, 5, 20\}$  & Number of covariates \\
    $\cv \in \{.1, 1\}$  & Heteroscedasticity metric (low, high) \\
    $\rng = 10$  &  Survival range metric \\
    $\rsq = .75$  & Proportion of variance explained \\
    $P_C \in \{0, .2, .5\}$  & Censoring probability (none, moderate, high) \\
    \bottomrule
  \end{tabular}
  \caption{Parameter values in the simulation study.}\label{tab:sim}
\end{table}
AFT and HAFT models were fit to each dataset using our accompanying software package\footnote{Details and a link to be provided upon acceptance of the manuscript.}.  The most challenging setting ($N = 1000$, $p = 20$, $P_C = 50\%$) took on average 50ms to fit on a personal computer.

In order to compare the performance of AFT and HAFT models for quantile estimation, the following statistics are recorded:
% For each experimental combination, we simulated $B = 1000$ datasets, $\DD_b = (\YY\sp b, \dde\sp b, \XX \sp b)$, $b = 1, \ldots, B$.  For the purpose of quantile estimation, the recorded statistics for each dataset are:
\begin{enumerate}
\item Noting that under the simulation model~\eqref{eq:simmodel}, the true conditional quantile
  \[
    Q_\alpha(\XX, \tth) = \exp\left\{\beta(1 + \bar X) + \sigma \exp(\gamma \bar X/2) \cdot \Phi^{-1}(\alpha)\right\}
  \]
  depends only on $\XX$ through $\bar X = \tfrac 1 p \sum_{j=1}^p X_j$, we record the normalized conditional root mean square error (RMSE)
  \[
    \rmse(\alpha, \bar X) = \frac 1 B \sum_{b=1}^BE\left[\frac{\big(\hat Q_\alpha\sp b(\XX) - Q_\alpha(\bar X)\big)^2}{Q_\alpha(\bar X)} \mid \bar X, \hat \tth\sp b\right],
  \]
  % \[
  %   \tx{Bias}\sp b(\alpha) = E\left[\re_\alpha(\XX, \hat \tth \sp b, \tth) \mid \hat \tth\sp b\right],
  % \]
  where $\hat Q_\alpha\sp b(\XX) = Q_\alpha(\XX, \hat \tth \sp b)$, $Q_\alpha(\bar X) = Q_\alpha(\bar X, \tth)$, and the conditional expectation is approximated numerically by sampling $M = 2000$ draws from the conditional distribution $p(\XX \mid \bar X)$.
  % \begin{align*}
  %   \re(\alpha; \XX, \hat \tth \sp b, \tth) & = \frac{Q_\alpha(\XX\sp b, \hat \tth\sp b) - Q_\alpha(\XX\sp b, \tth)}{Q_\alpha(\XX\sp b, \tth)} \\
  %   & = \exp\left\{\hat \beta_0 + \XX' \hat \bbe + \big(\hat \s e^{\XX'\hat \gga/2} + \s e^{\XX'\gga/2}\big) \cdot \Phi^{-1}(\alpha)\right\} - 1.
  % \end{align*}
  % \begin{align*}
  %   \tx{Bias}\sp b(\alpha)
  %   % & = \sum_{i=1}^N \frac{Q_\alpha(\XX_i\sp b, \hat \tth\sp b) - Q_\alpha(\XX_i\sp b, \tth)}{Q_\alpha(\XX_i\sp b, \tth)}
  %   & = E\left[\re(\alpha; \XX, \hat \tth \sp b, \tth) \mid \hat \tth\sp b\right] \\
  %   & = E\left[\exp\left\{\hat \beta_0 + \XX' \hat \bbe + \big(\hat \s e^{\XX'\hat \gga/2} + \s e^{\XX'\gga/2}\big) \cdot \Phi^{-1}(\alpha)\right\} \mid \tth\sp b\right].
  % \end{align*}
\item We also record the true coverage of the asymptotic 95\% confidence intervals constructed via~\eqref{eq:qse}, namely
  \[
    \tx{Cover}(\alpha, \bar X) = \frac 1 B \sum_{b=1}^B \Pr\left(Q_\alpha(\bar X) \in \hat Q_\alpha\sp b(\XX) \pm 1.96 \cdot \se\big(\hat Q_\alpha\sp b(\XX)\big) \mid \bar X, \hat \tth \sp b\right),
  \]
  where the conditional probability is again approximated by averaging over $M = 2000$ Monte Carlo draws from $p(\XX \mid \bar X)$.
\end{enumerate}

\begin{figure}[!htb]
  \centering
  \includegraphics[width=.97\textwidth]{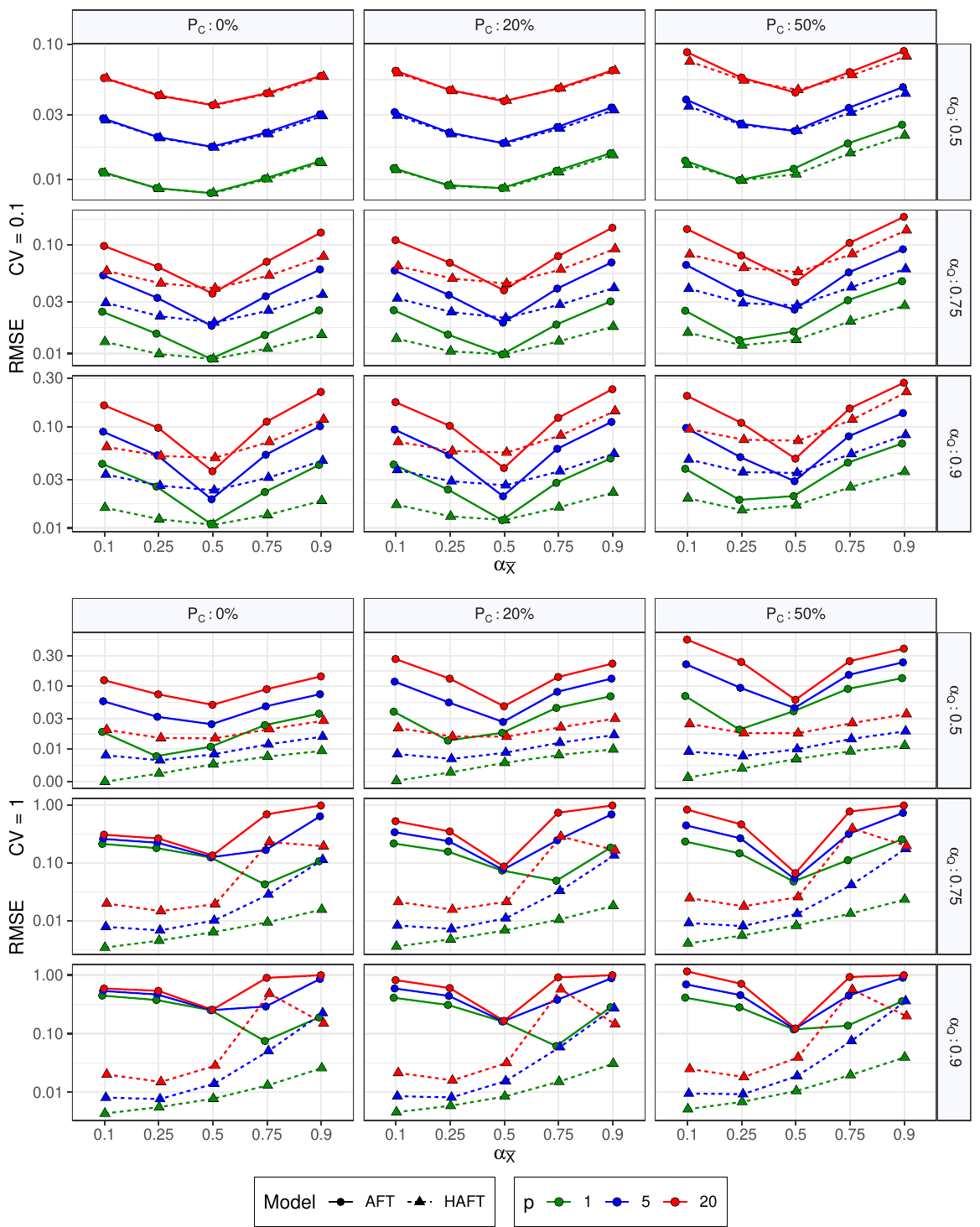}
  \caption{Normalized RMSE for conditional quantiles $Q_\alpha(\bar X)$ at $\alpha = \alpha_Q$ and $\bar X = \Phi^{-1}(\alpha_{\bar X})$.}
  \label{fig:sim_rmse}
\end{figure}
\begin{figure}[!htb]
  \centering
  \includegraphics[width=.97\textwidth]{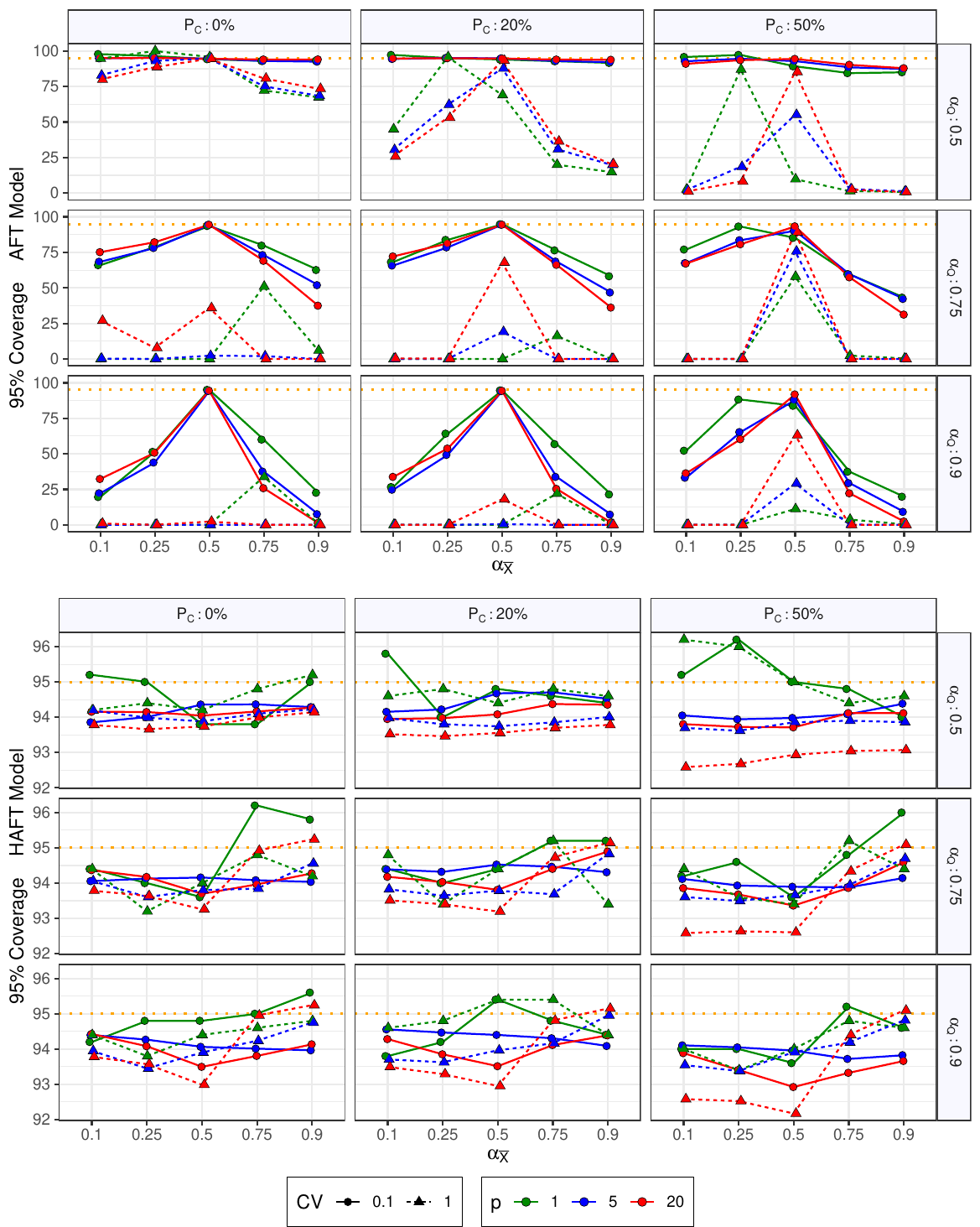}
  \caption{True coverage of 95\% confidence intervals for conditional quantiles $Q_\alpha(\bar X)$ at $\alpha = \alpha_Q$ and $\bar X = \Phi^{-1}(\alpha_{\bar X})$.  The dotted horizontal line indicates the 95\% nominal value.}
  \label{fig:sim_cover}
\end{figure}

Figure~\ref{fig:sim_rmse} displays the normalized $\rmse(\alpha, \bar X)$ over the experimental conditions in Table~\ref{tab:sim}, at various levels of the conditional quantile $\alpha = \alpha_Q$ and of the mean covariate level $\alpha_{\bar X}$, such that $\bar X = \Phi^{-1}(\alpha_{\bar X})$.  When the degree of heteroscedasticity is low ($\cv = 0.1$), both models have normalized RMSEs typically below 5-10\%.  However, heteroscedastic modeling becomes considerably more beneficial when $\cv = 1$, in which case normalized RMSEs under the homoscedasticity assumption jump to 50-100\% at the upper quantiles $\alpha_Q \in \{0.75, 0.9\}$ as $\bar X$ gets further from its median value.

Figure~\ref{fig:sim_cover} displays the true coverage probabilities of the 95\% confidence intervals given by~\eqref{eq:qse}.  In this case, the impact of failing to account for even a small degree of heteroscedasticity ($\cv = 0.1$) is much more severe,
% the impact of ignoring heteroscedasticity is much more severe,
as coverage falls well below the nominal 95\% level at the upper quantiles $\alpha_Q > 0.5$ when $\bar X$ is not at its median value.
% , even for $\cv = 0.1$.
In contrast, the HAFT confidence intervals achieve close to nominal coverage across the board, even for the challenging setting of 42 unknown parameters ($p = 20$ for $\bbe$, $\gga$, $\beta_0$ and $\sigma$) with 50\% censoring.

\section{Application to a Colon Cancer Study}\label{sec:eda}

The study of~\cite{laurie.etal89} and~\cite{moertel.etal90} is one of the first successful clinical trials of adjuvant chemotherapy for colon cancer.  Their dataset contains $N = 888$ patients with colon carcinoma randomly assigned to the control group (no treatment) or one of two chemotherapy treatment groups: levamisole combined with fluorouracil or levamisole alone.  In addition to the the treatment group, 10 covariates for each subject were also recorded (see Table~\ref{tab:data}).  Over half the survival times in the sample were right-censored ($N_\tx{cens} = 458$).
\begin{table}[!htb]
  \footnotesize
  \centering
  \begin{tabular}{rl}  
    \toprule
    Name & Description \\
    \midrule
    \tt{rx}  &  Treatment type: Control, Levamisole, Levamisole + Fluorouracil \\
    \tt{sex}  & Sex of patient \\
    \tt{age}  & Age of patient (in years) \\
    \tt{obstruct}  &  Obstruction of colon by tumor  (T/F) \\
    \tt{perfor}  &    Perforation of colon (T/F) \\
    \tt{adhere}  &    Adherence of tumor to nearby organs (T/F) \\
    \tt{nodes}  &     Number of lymph nodes with detectable cancer \\
    \tt{differ}  &    Differentiation index of tumor (1=well, 2=moderate, 3=poor)  \\
    \tt{extent}  &    Extent of local spread \\
         & (1=submucosa, 2=muscle, 3=serosa, 4=contiguous structures) \\
    \tt{surg}  &      Time from surgery to registration (0=short, 1=long)  \\
    \tt{node4}  &     More than 4 positive lymph nodes (T/F) \\
    \bottomrule
  \end{tabular}
  \caption{Description of covariates in the colon cancer trial.}\label{tab:data}
\end{table}

The purpose of this analysis is to estimate the conditional survival times of colon cancer patients given the predictors in Table~\ref{tab:data}. As a basis of comparison to the proposed HAFT model, we considered (i) the CPH model, (ii) a homoscedastic AFT model with log-normal survival times, and (iii) a linear quantile regression (QR) model
\[
  \Pr(T > \xx'\bbe_{\tau} \mid \XX = \xx) = \tau.
\]
The CPH and AFT models were fit with the \R package \tt{survival}~\citep{therneau15}, whereas the QR model was fit with the \R package \tt{quantreg}~\cite{koenker18}.  For quantile regression with censoring, both the estimators of \cite{portnoy03} and \cite{peng.huang08} were employed, corresponding to covariate-dependent generalizations of the Kaplan-Meier and Nelson-Aalen survival estimators, respectively~\citep{koenker08}.% -- using the \R package \tt{quantreg}~\citep{koenker18}.

In order to obtain well-fitting models, bidirectional stepwise regression based on the Akaike Information Criterion (AIC) was used to select the covariates in the AFT and CPH models amongst all main effects and second order interactions.  The HAFT model was given the same location covariates $\WW$ as its homoscedastic counterpart, followed by stepwise selection for the scale covariates $\ZZ$.  For this particular dataset, the only scale predictor retained is the treatment indicator \tt{rx}.  Full parameter estimates for the fitted models are given in Table~\ref{tab:coef}.
\begin{table}[!htb]
  \footnotesize
  \centering
  \begin{tabular}{lrrrrrrrr}
    \toprule
    \multicolumn{1}{c}{ } & \multicolumn{2}{c}{CPH} & \multicolumn{2}{c}{AFT} & \multicolumn{4}{c}{HAFT} \\
    \cmidrule(l{3pt}r{3pt}){2-3} \cmidrule(l{3pt}r{3pt}){4-5} \cmidrule(l{3pt}r{3pt}){6-9}
                          & $\hat \beta$ & $\se(\hat \beta)$ & $\hat \beta$ & $\se(\hat \beta)$ & $\hat \beta$ & $\se(\hat \beta)$ & $\hat \gamma$ & $\se(\hat \gamma)$\\
    \midrule
    \tt{(Intercept)} & . & . & 10.48 & 0.89 & 10.77 & 0.88 & 0.11 & 0.12\\
    \tt{rx(Lev)} & -0.24 & 0.19 & 0.07 & 0.17 & 0.15 & 0.17 & 0.31 & 0.18\\
    \tt{rx[Lev+5FU]} & -0.19 & 0.18 & 0.04 & 0.16 & 0.26 & 0.18 & 0.69 & 0.19\\
    \tt{sex[male]} & -1.04 & 0.52 & 0.94 & 0.50 & 0.83 & 0.49 & . & .\\
    \tt{age} & 0.03 & 0.01 & -0.02 & 0.01 & -0.03 & 0.01 & . & .\\
    \tt{obstruct} & 0.09 & 0.19 & -0.44 & 0.12 & -0.41 & 0.12 & . & .\\
    \tt{perfor} & 0.33 & 0.31 & -0.21 & 0.33 & -0.22 & 0.32 & . & .\\
    \tt{adhere} & 0.51 & 0.20 & -1.30 & 0.82 & -1.05 & 0.79 & . & .\\
    \tt{nodes} & 0.14 & 0.04 & -0.15 & 0.04 & -0.17 & 0.04 & . & .\\
    \tt{differ[moderate]} & 1.24 & 0.94 & -0.74 & 0.81 & -1.04 & 0.82 & . & .\\
    \tt{differ[poor]} & 3.55 & 1.01 & -2.55 & 0.93 & -2.94 & 0.93 & . & .\\
    \tt{extent[muscle]} & 0.39 & 0.61 & -0.24 & 0.45 & -0.28 & 0.45 & . & .\\
    \tt{extent[serosa]} & 0.91 & 0.59 & -0.79 & 0.43 & -0.80 & 0.43 & . & .\\
    \tt{extent[cstruct]} & 1.28 & 0.62 & -1.25 & 0.48 & -1.21 & 0.48 & . & .\\
    \tt{surg[long]} & 0.21 & 0.11 & -0.24 & 0.11 & -0.23 & 0.10 & . & .\\
    \tt{node4} & 0.48 & 0.19 & -0.44 & 0.19 & -0.36 & 0.19 & . & .\\
    \tt{obstruct:perfor} & -1.19 & 0.61 & 1.19 & 0.59 & 1.07 & 0.57 & . & .\\
    \tt{age:differ[moderate]} & -0.02 & 0.01 & 0.02 & 0.01 & 0.02 & 0.01 & . & .\\
    \tt{age:differ[poor]} & -0.06 & 0.02 & 0.04 & 0.01 & 0.04 & 0.01 & . & .\\
    \tt{age:sex[male]} & 0.02 & 0.01 & -0.02 & 0.01 & -0.02 & 0.01 & . & .\\
    \tt{rx(Lev):sex[male]} & 0.10 & 0.23 & -0.13 & 0.23 & -0.14 & 0.21 & . & .\\
    \tt{rx[Lev+5FU]:sex[male]} & -0.44 & 0.25 & 0.39 & 0.24 & 0.42 & 0.25 & . & .\\
    \tt{rx(Lev):obstruct} & 0.61 & 0.28 & . & . & . & . & . & .\\
    \tt{rx[Lev+5FU]:obstruct} & 0.04 & 0.31 & . & . & . & . & . & .\\
    \tt{adhere:nodes} & -0.06 & 0.03 & . & . & . & . & . & .\\
    \tt{adhere:age} & . & . & 0.02 & 0.01 & 0.01 & 0.01 & . & .\\
    \tt{adhere:differ[moderate]} & . & . & -0.12 & 0.53 & -0.20 & 0.51 & . & .\\
    \tt{adhere:differ[poor]} & . & . & 0.57 & 0.58 & 0.44 & 0.56 & . & .\\
    \bottomrule
  \end{tabular}
  \caption{Coefficient estimates and their standard errors for the CPH, AFT, and HAFT models.}\label{tab:coef}
\end{table}

\begin{figure}[!htb]
  \includegraphics[width=1.00\textwidth]{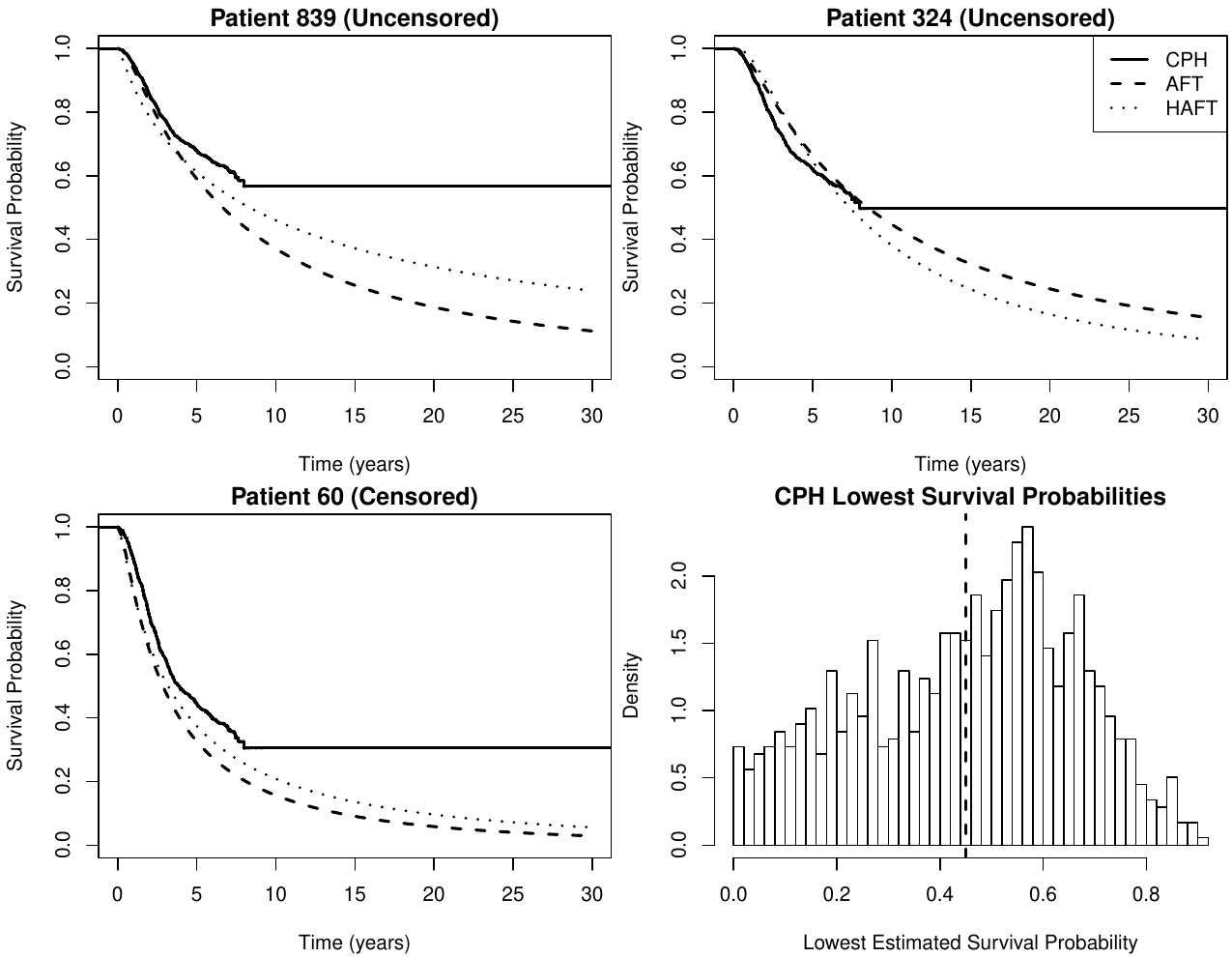}
  \caption{(a-c) Estimated survival curves for three randomly selected patients.  (d) Lowest estimated survivals for the CPH model (median indicated by dashed line).}
  \label{fig:cphsurv}
\end{figure}

Figures~\ref{fig:cphsurv}(a-c) display the estimated survival curves for the CPH, AFT, and HAFT models for three randomly selected subjects.  All models are in close agreement with each other at the lower tail of the conditional survival distribution, where the data is most informative.  However, due to the high proportion of censored observations, %the semiparametric CPH model does not produce estimates for the upper survival quantiles.  Indeed
the CPH model truncates more than half the estimated conditional survival curves below 50\% survival (Figure~\ref{fig:cphsurv}d).

As expected, the QR estimators of \citeauthor{portnoy03} and \citeauthor{peng.huang08} are undefined in the upper tail. %could not produce conditional quantile estimates in the upper tail.
However, in this heavy-censoring scenario the limitations were particularly severe: %PQR and PHQR
% could not estimate conditional quantiles b
the maximum quantile estimates available were 44\% and 41\% for the \citeauthor{portnoy03} and \citeauthor{peng.huang08} methods, respectively, 
% could not estimate conditional survivals beyond the lowest 44\% and 41\% quantiles, respectively,
for a model with only the marginal treatment effect \tt{rx},
and only 16\% and 14\%, respectively, for a model with only the main covariate effects.
%and not beyond the lowest 16\% and 14\% quantiles with only the main covariate effects.  %Therefore, we drop the quantile regression estimators from the remainder of our analysis.
As the semiparametric CPH and QR models were thus deemed ill-suited to estimate conditional quantiles for this particular dataset, we focus on the parametric AFT and HAFT models for the remainder of the analysis.  

% To pick representatives of the CPH and homoscedastic AFT families of models, stepwise selection with the Akaike Information Criterion (AIC) was used

% Stepwise regression based on the AIC was employed to select the covariates in the AFT and CPH models amongst all main effects and second order interactions.  The HAFT model was given the same location covariates as its homoscedastic counterpart, and for simplicity we set the shape covariates to all the main effects.  Parameter estimates for the fitted models are in Appendix~\ref{app:modelfit}.

\subsection{Goodness-of-Fit Residual Analysis}\label{sec:good}

The AIC statistics for the AFT and HAFT models are 2022.1 and 2012.8, thus distinctly favoring the HAFT model. %(the AIC for the CPH model is calculated from a partial likelihood and thus cannot be compared directly to the other two).
To further compare these models, a goodness-of-fit analysis based on the following definition of model residuals is proposed.

% to its heteroscedastic extension, we consider the following goodness-of-fit tests for the model residuals.

For a given parametric conditional survival model $p(R \mid \XX, \tth)$, we would like to compare the log-survival time $R_i$ of each patient to its predictive distribution $p(R_i \mid \XX_i, \hat {\tth})$.  In the absence of censoring, the HAFT model residuals are
\[
  \hat \veps_i = \frac{R_i - \WW_i'\hat {\bbe}}{\exp(\ZZ_i'\hat {\gga}/2)}.
\]

With censoring, however, we do not observe $R_i$ but instead $(Y_i, \dd_i)$, with $Y_i = \min(R_i, C_i)$ and $\delta_i = \mathfrak 1\{R_i < C_i\}$.  A common approach to defining model residuals in the presence of censoring is to impute the missing survivals times~\citep{hillis95}.  That is, each censored observation is given a stochastic residual $\tilde \veps_i$, computed as above but with $\tilde R_i$ drawn from its truncated conditional distribution,
\[
  \tilde R_i \sim p(R \mid R > Y_i, \XX_i, \hat {\tth}).
\]
The resulting Hillis residuals are approximately $\N(0,1)$ under a correctly specified HAFT model.  However, in the presence of heavy censoring, %as in the colon cancer study,
the Hillis residuals which are simulated from the posited model can easily overwhelm the uncensored data, and thus significantly decrease the power of goodness-of-fit tests.

Instead, we propose to construct more stringent model residuals by parametrically modeling both the conditional survival \emph{and} censoring distributions.  
% by first fitting a second parametric model, now to the conditional distribution of censoring times.
While this requires additional assumptions, the large number of censored observations provided sufficient information to select AFT and HAFT candidate models for $p(C \mid \XX)$, exactly as for the survival distribution but with status indictor $1-\delta$.

Let $f_{R\mid \XX}(r \mid \xx)$, $F_{R\mid\XX}(r \mid \xx)$ and $f_{C\mid\XX}(c \mid \xx)$, $F_{C\mid\XX}(c \mid \xx)$ denote the condition PDF and CDF of survival and censoring distributions, respectively.  Then the conditional PDF of the observed survival time $Y = \min(R, C)$ is
\begin{equation}\label{eq:obscond}
\begin{aligned}
  f_{Y\mid\dd,\XX}(y \mid \delta = 1, \XX = \xx) & \propto f_{R\mid\XX}(y \mid \xx)\cdot \Big(1 - F_{C\mid\XX}(y \mid \xx)\Big), \\%\label{eq:obssur}\\
  f_{Y\mid\dd,\XX}(y \mid \delta = 0, \XX = \xx) & \propto f_{C\mid\XX}(y \mid \xx)\cdot \Big(1 - F_{R\mid\XX}(y \mid \xx)\Big), % \label{eq:obscens}
\end{aligned}
\end{equation}
for uncensored and censored observations, respectively.

While the conditional distributions for the AFT and HAFT models are normal, the conditional distributions in~\eqref{eq:obscond}
%~\eqref{eq:obssur} and~\eqref{eq:obscens}
are not.  Our stringent model residuals are constructed by mapping each observation $Y_i$ to its predicted normal quantile:
\begin{equation}\label{eq:resid}
  \hat \veps_i := \Phi^{-1}\Big(P(Y \le Y_i \mid \dd_i, \XX_i, \hat {\tth})\Big),
\end{equation}
where $P(Y \le y \mid \dd, \XX, \tth)$ is the CDF associated with the PDFs in~\eqref{eq:obscond}.% and~\eqref{eq:obscens}.
% , and $\Phi(\cdot)$ is the CDF of the standard normal distribution.
The inner term in~\eqref{eq:resid} thus corresponds to the probability integral transform of $Y_i$, such that the $\hat \veps_i$ are approximately standard normal when both the survival and censoring models are correctly chosen.  The residuals in~\eqref{eq:resid} are more stringent than those of Hillis, not only because they avoid simulating data which artificially improves the goodness of fit, but also by exacting that both conditional survival and censoring models be specified correctly.

\subsubsection{Results} \label{sec:rslt}

\begin{figure}[!htb]
  \centering
  \includegraphics[width=\textwidth]{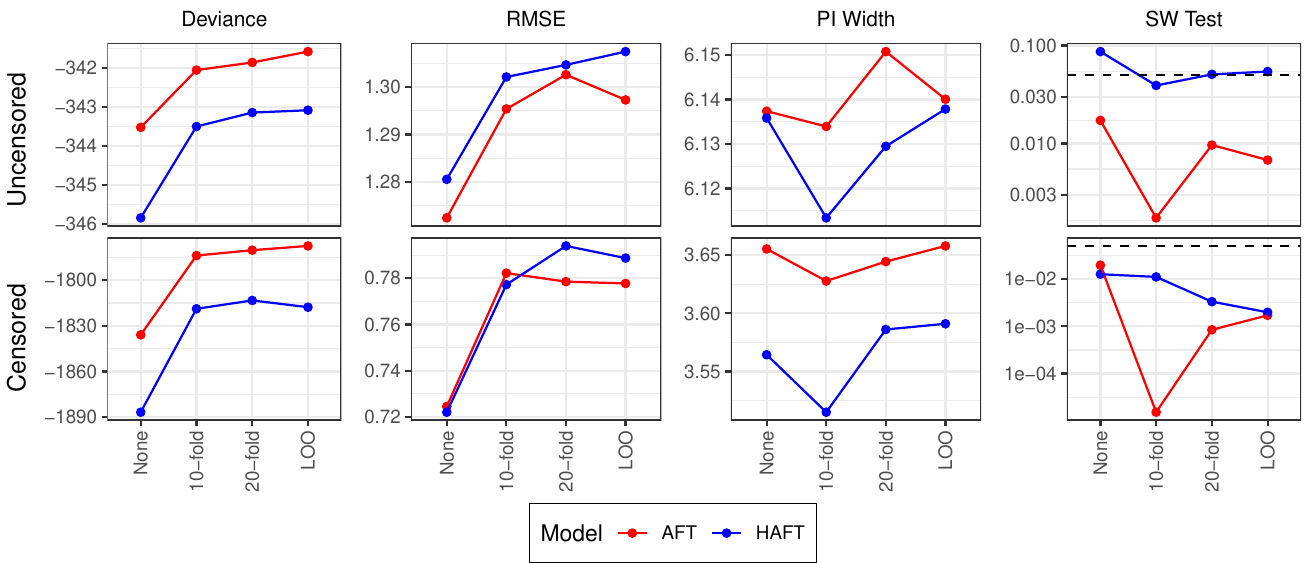}
  \caption{Various goodness-of-fit statistics for AFT and HAFT models grouped by censoring status and type of cross-validation (None: no cross-validation, LOO: leave-one-out). RMSE and PI Width are in units of years.  The nominal 5\% p-value cutoff for the SW Test is given by the dotted line.}
  \label{fig:fitstats}
\end{figure}

Figure~\ref{fig:fitstats} displays various goodness-of-fit metrics comparing the AFT and HAFT models based on the estimated observed lifetime distribution $p(Y \mid \delta, \XX)$ defined by~\eqref{eq:obscond}.  Each model is thus fitted twice, in order to estimate both the conditional survival and censoring distributions $p(R \mid \XX)$ and $p(C \mid \XX)$.
% residuals in~\eqref{eq:resid}.  Each model is used to estimate both the conditional survival and censoring distributions, $p(R \mid \XX)$ and $p(C \mid \XX)$, in order to construct the 
% For each model we estimate both conditional survival and censoring distribution, $p\sp m(R \mid \XX)$ and $p\sp m(C \mid \XX)$, $m \in \{\tx{AFT}, \tx{HAFT}\}$, in order to construct the conditional observed lifetime distribution $p\sp m(Y \mid \delta, \XX)$.  In addition
The first column of Figure~\ref{fig:fitstats} displays the deviance statistic,
\[
  D^{\Mo}_{\St} = -2 \sum_{i \in \St}\log p(Y_i \mid \delta_i, \XX_i, \hat \tth_\Mo),
\]
where $\Mo \in \{\tx{AFT}, \tx{HAFT}\}$ and $\St$ is the set of either the uncensored or the censored observations.  In addition to the calculating the deviance on the whole dataset, we calculate its average value over 10-fold, 20-fold, and leave-one-out (LOO) cross-validation settings.  The results are fairly close for the uncensored observations, but for the censored observations they distinctly favor HAFT.

The second column of Figure~\ref{fig:fitstats} displays the root mean square error (RMSE)
\[
  \tx{RMSE}_{\St}^{\Mo} = \bsqrt{\frac 1 {|\St|} \sum_{i\in \St} \big(\exp(Y_i) - E[\exp(Y_i) \mid \delta_i, \XX_i, \hat \tth_\Mo]\big)^2}
\]
under the same conditions as above (in units of years).  In this case, the AFT model performs slightly better, although the largest difference (Uncensored LOO) is on the order of about five days.

The third column of Figure~\ref{fig:fitstats} displays the average width of the 95\% prediction intervals
\[
  \tx{PI}_{\St}^{\Mo} = \frac 1{|\St|} \sum_{i\in \St} Q_{.975}(Y_i \mid \delta_i, \XX_i, \hat \tth_\Mo) - Q_{.025}(Y_i \mid \delta_i, \XX_i, \hat \tth_\Mo).
\]
As expected, the richer HAFT model has narrower prediction intervals, although the difference is very small (at most about 25 days).

\begin{figure}[!htb]
  \centering
  \includegraphics[width=\textwidth]{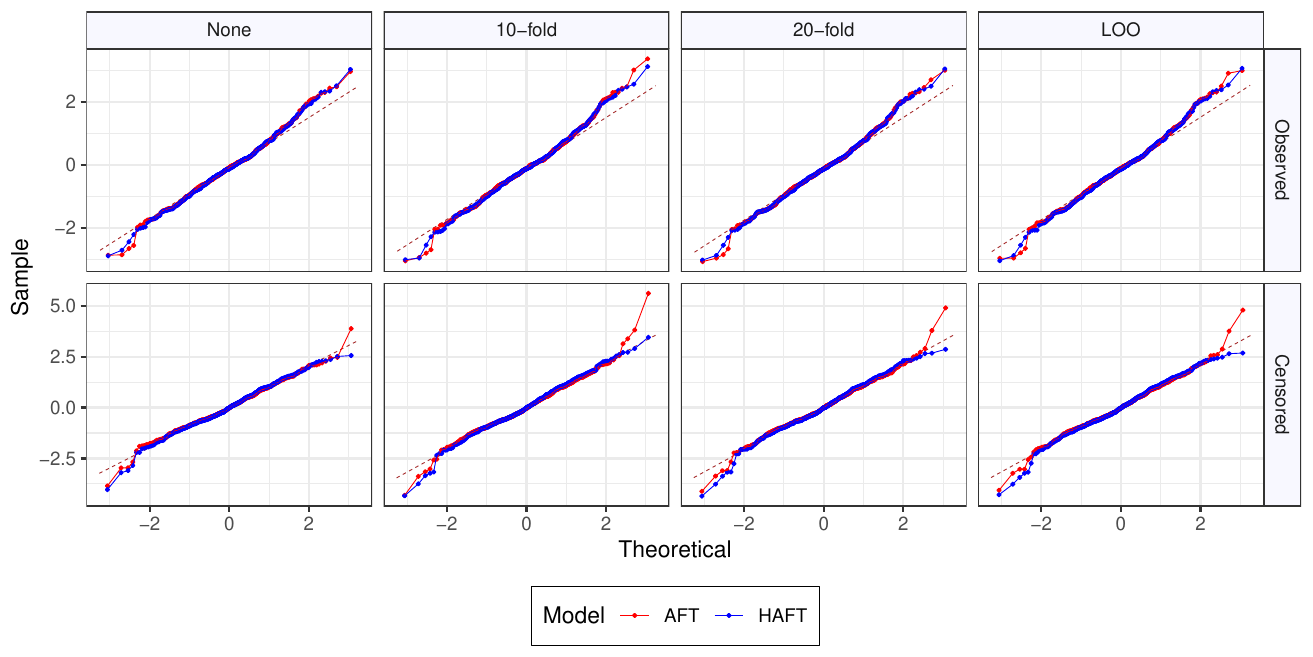}
  \caption{QQ-plots of the normalized observed residuals $\hat \veps_i$~\eqref{eq:resid} for AFT and HAFT models, grouped by censoring status and type of cross-validation (None: no cross-validation, LOO: leave-one-out).}
  \label{fig:fitqq}
\end{figure}
The final column of Figure~\ref{fig:fitstats} displays the p-value of the Shapiro-Wilk normality test~\citep{shapiro.wilk65} on the normalized observed residuals $\hat \veps_i = \Phi^{-1}\Big(P(Y \le Y_i \mid \dd_i, \XX_i, \hat \tth_\Mo)\Big)$ defined by~\eqref{eq:resid}.  Given that the Shapiro-Wilk test is particularly powerful at detecting departures from the null~\citep[e.g.,][]{razali.wah11}, it is noteworthy that it does not reject normality at the 5\% level for the uncensored observations.  We explore this finding more carefully in the QQ-plots of Figure~\ref{fig:fitqq}, which reveal that HAFT removes many of the extreme AFT residuals found in the upper tail.

\subsection{Quantile Estimation}

\begin{figure}[!htb]
  \centering
  \includegraphics[width=\textwidth]{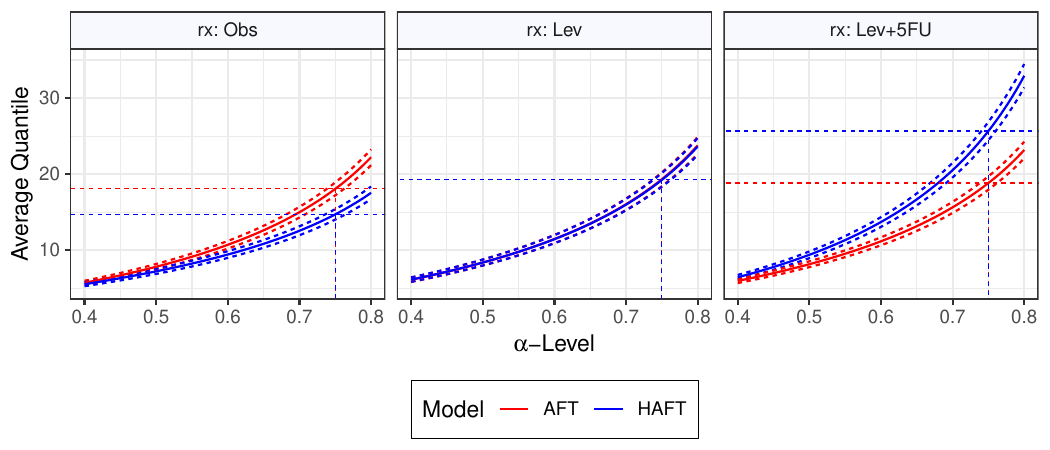}
  \caption{Predicted mean population quantile $\bar Q(\alpha, \tt{rx})$ for AFT and HAFT models at vairous quantile levels $\alpha$ and treatment $\tt{rx}$ (\tt{Obs}: Controls, \tt{Lev}: Levamisole, \tt{Lev+5FU}: Levamisole and Fluorouracil).  The 75\% quantiles are indicated by the dashed lines.}
  \label{fig:fitqpred}
\end{figure}

We now address the stated purpose of estimating the conditional survival times of the colon cancer patients.  The quantity of interest is defined as the mean population quantile at a given level $\alpha$ and treatment \tt{rx}:
\begin{equation}\label{eq:qpop}
  \bar Q(\alpha, \tt{rx}) := E[Q_T(\alpha \mid \tt{rx}, \tilde \XX)],
\end{equation}
where the expectation is over $\tilde \XX$, which corresponds to all covariates in Table~\ref{tab:data} except $\tt{rx}$.  Figure~\ref{fig:fitqpred} displays estimates of $\bar Q(\alpha, \tt{rx})$ for AFT and HAFT models, given by
%estimated mean population quantile
%average population quantile for AFT and HAFT models at various levels of $\alpha$ at each treatment level \tt{rx}.  That is, for a given treatment $\tt{rx} = k$ and quantile level $\alpha$, we calculate
\[
  \bar Q_{\Mo}(\alpha, k) = \frac 1 N \sum_{i=1}^N \exp\left\{\WW_i^{(k)\prime}\hat \bbe_\Mo + \exp(\ZZ_i^{(k)\prime}\hat \gga_\Mo/2) \cdot \Phi^{-1}(\alpha)\right\},
\]
where $(\hat \bbe_{\Mo}, \hat \gga_{\Mo})$ are MLEs calculated from the entire dataset, and $(\WW_i\sp k, \ZZ_i\sp k)$ are the location and scale covariates for individual $i$, but with the (possibly counterfactual) treatment value $\tt{rx} = k$.  %The parameter estimates for the AFT and HAFT model are obtained without cross-validation.
Both AFT and HAFT models predict a small improvement in the population quantile metric~\eqref{eq:qpop} due to levamisole (\tt{Lev}) or levamisole and fluorouracil (\tt{Lev+5FU}) treatment at the 50\% quantile level.  While this is still true of the AFT model at the 75\% quantile level, there the effect predicted by HAFT is much more pronounced, corresponding to a 10-year lifetime extension for \tt{Lev+5FU} treatment compared to the controls (\tt{Obs}).

\section{Discussion}\label{sec:disc}

The heteroscedastic AFT model proposed here is a natural extension to its homoscedastic counterpart, admitting tractable conditional survival quantile estimates in the presence of heavy right-censoring, which many nonparametric and semiparametric estimators fail to produce.
Results of a simulation study indicate that even a small degree of unaccounted heteroscedasticity can lead to severe bias and undercoverage of conditional quantile estimators.
% and benefits from tractable computations in the presence of right-censoring.
In an analysis of a colon cancer clinical trial, 
% study which features heavy censoring,
the HAFT model was found to exhibit substantially fewer outliers than the homoscedastic AFT, and predict better response to treatment in the upper-tail quantiles.
% while slightly decreasing the average size of prediction intervals.
% diminish the significance of model outliers, without increasing the average size of prediction intervals.

The results of this study are promising for the HAFT model, prompting several possible extensions to more complex models or with fewer assumptions.  For instance, the ECM algorithm in Section~\ref{sec:necm} could be adapted to heavy-tailed residuals %could be incorporated
via the t-distribution~\citep[e.g.,][]{arellano-valle.etal12}.  Alternatively, one might choose not to specify the residual distribution, in which case a number of semiparametric homoscedastic AFT models can be adapted to the heteroscedastic setting~\citep[e.g.,][]{buckley.james79,robins.tsiatis92,zhang.davidian08,zhou.etal12,daye.etal12}.  Similarly, it is possible to embed the HAFT model within more complex modeling frameworks to account for individual-level random effects or competing risks.  It is anticipated that the computational simplicity of the proposed HAFT model can be leveraged to design effective Monte Carlo inference strategies in these more sophisticated settings.

% \subsubsection*{Acknowledgements} This work is supported by the \nserc{} grant number \grantnumber.

% ------------------------------------------------

% ----------------------------------------------------------------------------------------
%	BIBLIOGRAPHY
% ----------------------------------------------------------------------------------------

% \newpage
\bibliographystyle{apalike}
\bibliography{hlm-ref}

\end{document}